\documentclass[apj]{emulateapj}
\usepackage{apjfonts}
\usepackage{graphicx}

\slugcomment{Accepted for Publication in the Astrophysical Journal Letters}

\shorttitle{Methanol Ice in the Central Molecular Zone}
\shortauthors{An et~al.}

\begin{document}

\title{Abundant Methanol Ice toward a Massive Young Stellar Object\\ in the Central Molecular Zone
\footnote{Based in part on data collected at Subaru Telescope, which is operated by the National Astronomical Observatory of Japan.}}

\author{Deokkeun An\altaffilmark{1,7}, Kris Sellgren\altaffilmark{2,7}, A. C. Adwin Boogert\altaffilmark{3},
Solange V. Ram{\'i}rez\altaffilmark{4}, Tae-Soo Pyo\altaffilmark{5,6}}

\altaffiltext{1}{Department of Science Education, Ewha Womans University, 52 Ewhayeodae-gil, Seodaemun-gu, Seoul 03760, Korea; deokkeun@ewha.ac.kr}
\altaffiltext{2}{Department of Astronomy, Ohio State University, 140 West 18th Avenue, Columbus, OH 43210, USA}
\altaffiltext{3}{Universities Space Research Association, Stratospheric Observatory for Infrared Astronomy, NASA Ames Research Center, MS 232-11, Moffett Field, CA 94035, USA}
\altaffiltext{4}{NASA Exoplanet Science Institute, California Institute of Technology, Mail Stop 100-22, Pasadena, CA 91125, USA}
\altaffiltext{5}{Subaru Telescope, National Astronomical Observatory of Japan, National Institutes of Natural Sciences (NINS), 650 North A'ohoku Place, Hilo, HI 96720, USA}
\altaffiltext{6}{School of Mathematical and Physical Science, SOKENDAI (The Graduate University for Advanced Studies), Hayama, Kanagawa 240-0193, Japan}
\altaffiltext{7}{Visiting Astronomer at the Infrared Telescope Facility, which is operated by the University of Hawaii under contract NNH14CK55B with the National Aeronautics and Space Administration.}

\begin{abstract}

Previous radio observations revealed widespread gas-phase methanol (CH$_3$OH) in the Central Molecular Zone (CMZ) at the Galactic center (GC), but its origin remains unclear. Here, we report the discovery of CH$_3$OH ice toward a star in the CMZ, based on a Subaru $3.4$--$4.0\ \mu$m spectrum, aided by NASA/IRTF $L'$ imaging and $2$--$4\ \mu$m spectra. The star lies $\sim8000$~au away in projection from a massive young stellar object (MYSO). Its observed high CH$_3$OH ice abundance ($17\%\pm3\%$ relative to H$_2$O ice) suggests that the $3.535\ \mu$m CH$_3$OH ice absorption likely arises in the MYSO's extended envelope. However, it is also possible that CH$_3$OH ice forms with a higher abundance in dense clouds within the CMZ, compared to within the disk. Either way, our result implies that gas-phase CH$_3$OH in the CMZ can be largely produced by desorption from icy grains. The high solid CH$_3$OH abundance confirms the prominent $15.4\ \mu$m shoulder absorption observed toward GC MYSOs arises from CO$_2$ ice mixed with CH$_3$OH.

\end{abstract}


\section{Introduction}

Methanol (CH$_3$OH) is a key species in the formation of complex organic molecules. CH$_3$OH is observed in gas phase throughout the Central Molecular Zone (CMZ), the innermost $\sim400$~pc region of the Milky Way \citep[see][]{morris:96}, with a wide variation in abundance relative to H$_2$ \citep[$\sim10^{-8}$--$10^{-6}$;][]{requena-torres:06,yusefzadeh:13}. Complex organic molecules are also observed throughout the CMZ, with high and constant gas-phase abundances, relative to gas-phase CH$_3$OH, that are matched by hot cores in the Galactic disk \citep{requena-torres:06}.

Gas-phase production of CH$_3$OH is inefficient, but formation by the hydrogenation of CO ices in dense ($\sim10^5$\,cm$^{-3}$) and cold ($\sim15$~K) environments, such as dense molecular cores, is very active as shown by laboratory experiments and Monte Carlo simulations \citep[e.g.,][]{watanabe:03,cuppen:09,coutens:17}. A number of non-thermal desorption mechanisms, including shocks \citep{requena-torres:08}, cosmic rays \citep{yusefzadeh:13}, and episodic explosions \citep{coutens:17} have been proposed to explain the widespread gas-phase CH$_3$OH \citep[possibly along with other complex molecules; see][]{rawlings:13} in the CMZ. However, previous searches for CH$_3$OH ice in the CMZ have been unsuccessful; only upper limits on the abundance of solid CH$_3$OH relative to H$_2$O ice, $N_{\rm solid}$(CH$_3$OH)/$N_{\rm solid}$(H$_2$O), have been established from Sgr~A* ($<0.04$) and the Quintuplet cluster star \objectname[WR~102dd]{GCS 3-I} ($<0.27$) \citep{chiar:00,gibb:04,moultaka:15}.

Indirect evidence of the presence of solid CH$_3$OH came from mid-infrared spectra of massive young stellar objects (MYSOs) in the Galactic center (GC) region \citep{an:09,an:11}. We identified these GC MYSOs by their wide absorption profile of $15\ \mu$m CO$_2$ ice, with a strong `shoulder' absorption component centered at $15.4\ \mu$m. To date, this 15.4 $\mu$m shoulder has only been observed toward disk MYSOs \citep[][and references therein]{boogert:15}. Laboratory studies attribute the $15.4\ \mu$m shoulder absorption to Lewis interaction of CO$_2$ in icy grains with other molecules such as methanol, ethanol, butanol, or diethylether \citep{dartois:99a}. CH$_3$OH has the highest abundance of these toward Galactic disk MYSOs.

The question emerges as to whether the observed $15.4\ \mu$m shoulder CO$_2$ ice absorption in the CMZ is produced by Lewis-base molecules other than CH$_3$OH because of the unusual CMZ conditions. CMZ molecular clouds are warmer, denser, and more turbulent than molecular clouds in the disk. Stronger tidal shear forces and magnetic fields pervade the CMZ \citep[see][and references therein]{morris:96}. The cosmic ray ionization rate is higher than in the disk \citep{goto:08}.  All these effects may complicate and lead to chemical networks that are distinct from those under local cloud conditions.

In this letter, we report the first direct detection of solid CH$_3$OH in the CMZ by searching for the $3.535\ \mu$m $\nu_3$ C--H stretching mode of CH$_3$OH ice in the GC MYSO \object[SSTGC~726327]{SSTGC~726327}. The $3.535\ \mu$m absorption is almost independent of ice mantle composition \citep{hudgins:93,ehrenfreund:99}, and therefore can be used to reliably measure a column density of solid CH$_3$OH. SSTGC~726327 is one of $35$ MYSOs identified in the GC using $5$--$35\ \mu$m {\it Spitzer}/IRS spectra \citep{an:09,an:11}, with both the $15.4\ \mu$m shoulder CO$_2$ ice and $14.97\ \mu$m gas-phase CO$_2$ absorption. Its silicate feature implies an integrated (both foreground and internal) visual extinction $A_V = 46\pm3$~mag, while the foreground extinction from the Galactic disk is $30\pm4$~mag \citep{schultheis:09}.

\section{Observations and Data Reductions}

\subsection{IRTF}

We acquired $L'$-band images of SSTGC~726327 with NSFCAM2 \citep{shure:94} at the 3.0~m NASA/IRTF on 2010 July 19 (UT). We took two sets of images in a $3\times3$ dither pattern, yielding a total effective exposure time of $315$~s for each set. The FWHM of image was $0.5\arcsec$. We used standard {\tt IRAF}\footnote{IRAF is distributed by the National Optical Astronomy Observatory, which is operated by the Association of Universities for Research in Astronomy (AURA) under a cooperative agreement with the National Science Foundation.} routines to reduce images and extract photometry. SSTGC~726327 is a point source in [3.6] IRAC images \citep{ramirez:08}. Our higher-resolution $L'$ image resolves SSTGC~726327 into faint extended emission plus two point sources separated by 2\arcsec: SSTGC~726327E  (\object{UGPS J174653.31-283201.2}) and SSTGC~726327W  (\object{UGPS J174653.16-283201.5}).

In addition to the $L'$ images, we obtained a $2$--$4\ \mu$m spectrum of SSTGC~726327E with SpeX \citep{rayner:03} at the NASA/IRTF, on 2009 May 15 (UT).  We observed through varying amounts of cirrus in the long cross-dispersed (LXD) mode. The slit width of $0.5\arcsec$, matching the seeing, resulted in $\lambda / \Delta \lambda$ = $R$ $\approx$ $1,500$.  We nodded the telescope along the north--south slit.  The total on-source integration time was $32$~minutes.

Flat field, argon arc, and telluric standard spectra were taken immediately before and after the observation. The airmass difference between the object and telluric standard spectra was negligible.  The spectra were extracted using {\tt Spextool}, a spectral extraction package for SpeX \citep{cushing:04}, in the point source extraction mode.  We corrected for telluric absorption following the standard procedures in \citet{vacca:03}. We flagged wavelengths with strong atmospheric absorptions from a telluric standard spectrum by taking a difference from its pseudo continuum, which was fit by a $5^{th}$-order polynomial in each spectral order. We removed data points with less than $50\%$ atmospheric transmission. The signal-to-noise ratio (S/N) of the final spectrum is $18$ at $3.6\ \mu$m--$4.0\ \mu$m.

\subsection{Subaru}

We obtained $3.4$--$4.0\ \mu$m high-resolution spectra of SSTGC~726327E and SSTGC~726327W with the echelle mode of IRCS \citep{kobayashi:00}, combined with an adaptive optics (AO) system, at the 8.2~m Subaru telescope on 2014 June 20 (UT).  The sky was mostly clear.  The seeing was $0.2\arcsec$ in FWHM after the AO correction. We nodded the two objects along the slit ($0.28\arcsec \times 6.69\arcsec$, $R\sim10,000$) for sky subtraction. We used LB$^-$, LB$^0$, and LB$^+$ configurations to have continuous spectral coverage.  The total on-source integration time for SSTGC~726327E was $3.1$ hr in LB$^0$, which covers most of the CH$_3$OH ice band, and $0.7$ hr each in LB$^-$ and LB$^+$. 

We observed a  set of A-type stars every hour for wavelength, telluric, and flux calibrations.  We grouped individual frames, depending on the time of observations and air masses, and made co-added echelle frames before extraction.  We extracted spectra using {\tt IRAF} routine {\tt apall} with an extraction aperture of $0.9\arcsec$. We corrected for hydrogen absorption lines in the standard star spectra using a smoothed theoretical spectrum of Vega \citep{castelli:04}\footnote{http://kurucz.harvard.edu/stars/vega/}, and then derived telluric corrections. We flagged and removed data where the atmospheric transmission, relative to an observed psuedo continuum, was $<50\%$. We averaged the flux into $3$~nm wide bins in steps of $1.5$~nm. Our final spectrum has $R \approx600$ and S/N of $42$ per binned data point in LB$^0$ and S/N$\approx24$ in LB$^-$ and LB$^+$. The SSTGC~726327W spectrum was extremely noisy and will not be discussed further.

Our flux calibrated Subaru/IRCS spectra are fainter than IRTF/SpeX $L$-band spectra by almost a factor of two, and show mild flux differences ($\sim2$--$3\%$) in order overlaps. To correct for this, we scaled the IRCS spectra to match our IRTF spectrum by taking a median difference in logarithmic flux at $3.6\ \mu$m--$3.9\ \mu$m for each of the three echelle configurations.

\section{Results}

\begin{figure}
\centering
\includegraphics[scale=0.45]{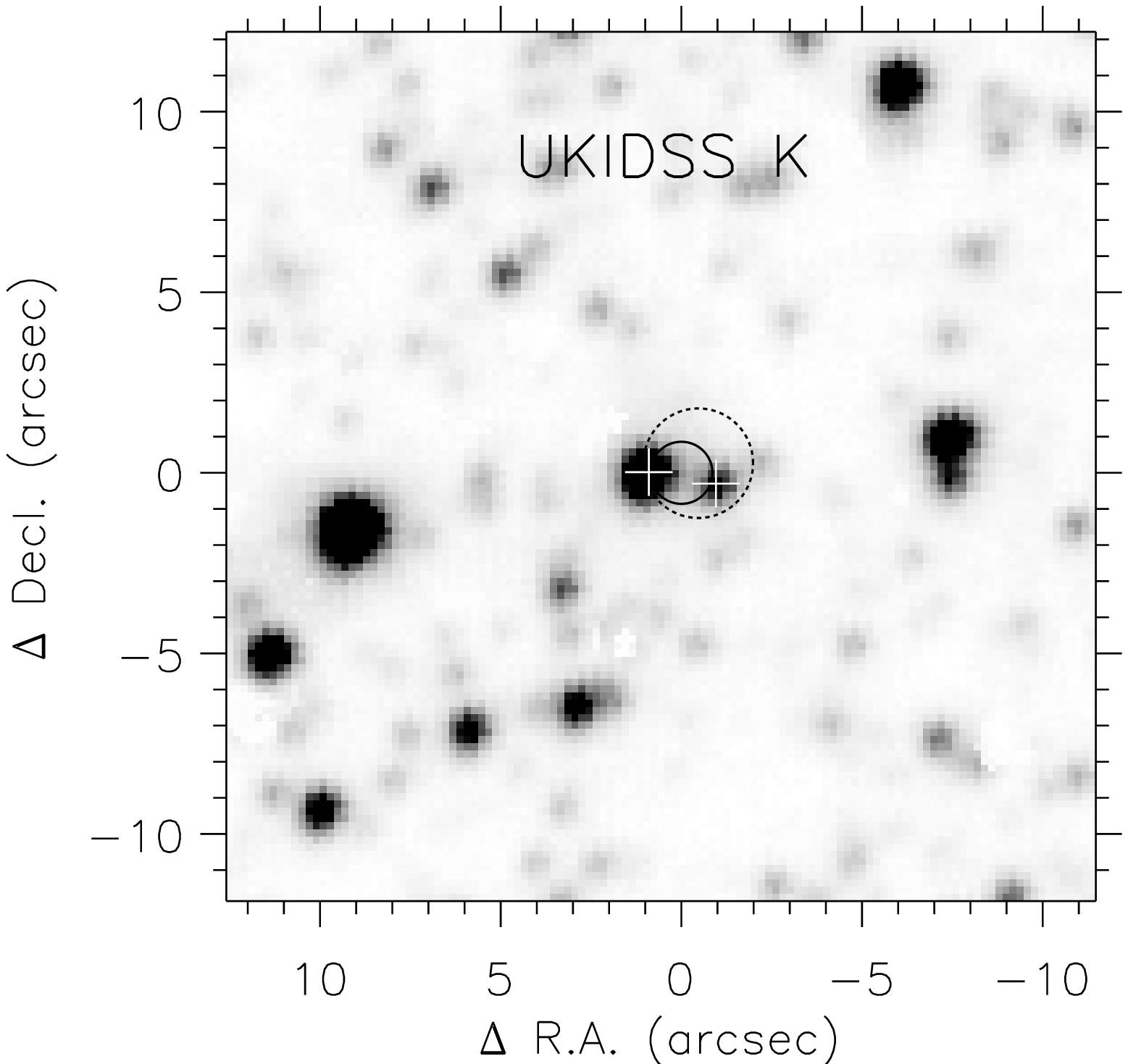}
\includegraphics[scale=0.45]{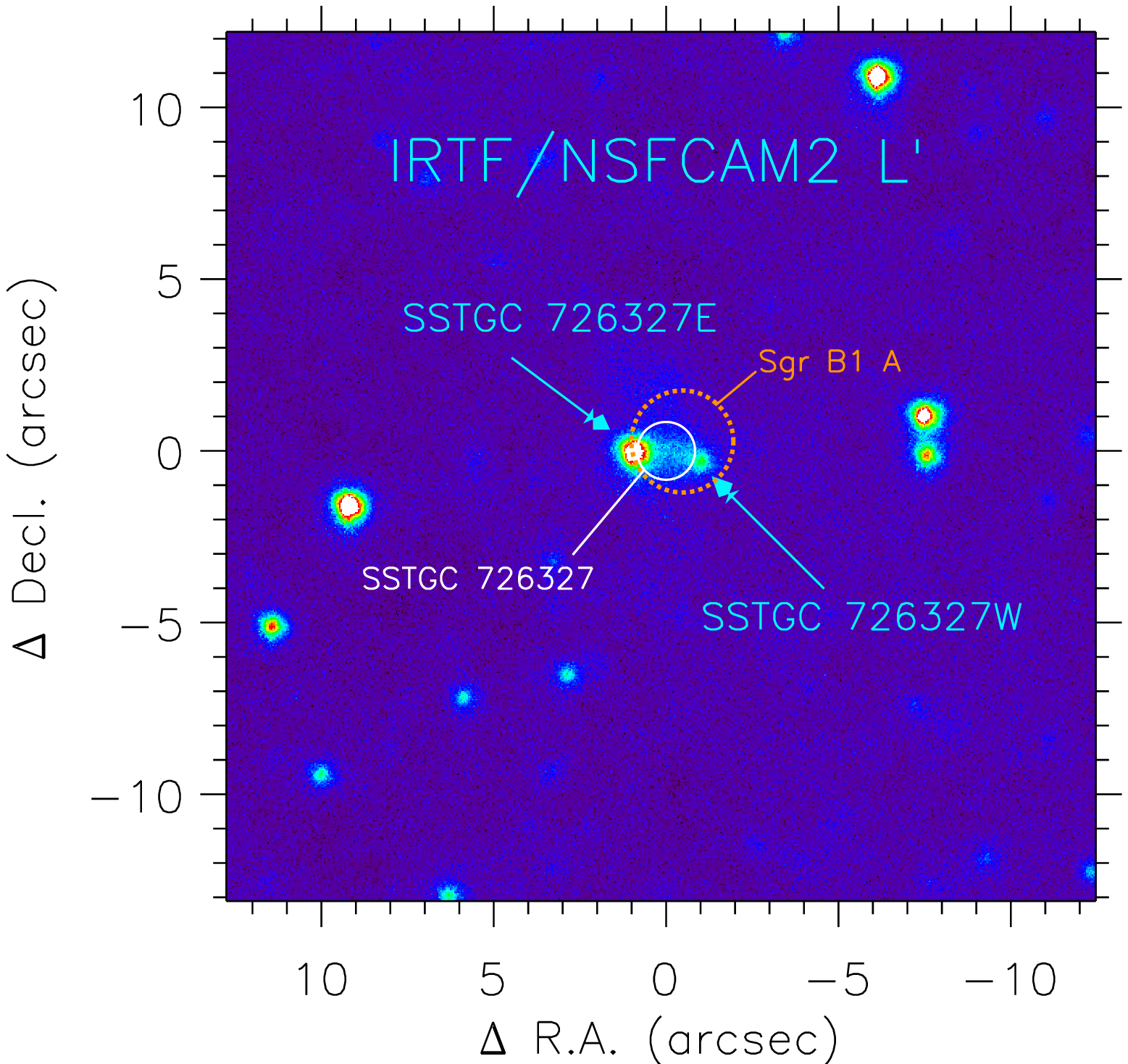}
\includegraphics[scale=0.45]{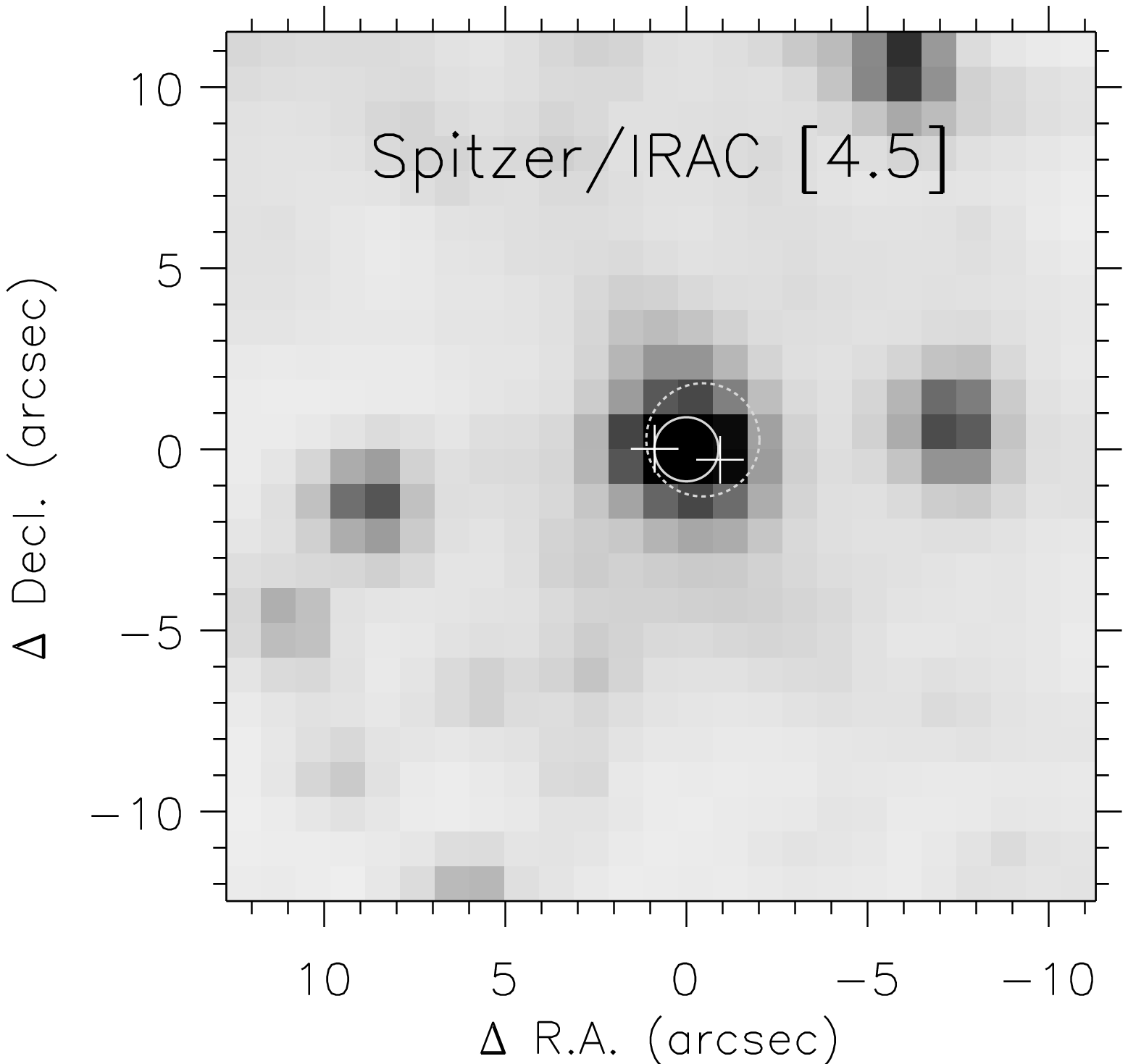}
\caption{Near- and mid-IR images centered on SSTGC~726327.  Top: $K$ image from the UKIDSS Galactic Plane Survey \citep[$0.8\arcsec$ resolution;][]{lucas:08}. Middle: $L'$ image from IRTF/NSFCAM2 ($0.5\arcsec$ resolution; this paper).  Bottom: {\it Spitzer}/IRAC [4.5] image \citep[$1.7\arcsec$ resolution;][]{ramirez:08}.  The $L'$ image shows that the MYSO SSTGC~726327 (solid circle with $1.7\arcsec$ diameter), a point source in {\it Spitzer}/IRAC images, is resolved into two point sources (SSTGC~726327E and SSTGC~726327W; plus symbols) plus diffuse emission. A dotted circle marks the location and diameter ($3\arcsec$) of the \ion{H}{2} region Sgr~B1~A \citep{mehringer:92}. North is to the top and east is to the left.} \label{fig:image} \end{figure}

Figure~\ref{fig:image} compares our $L'$ image (middle) with the $K$-band image from the UKIDSS Galactic Plane Survey \citep[][top]{lucas:08}, and the {\it Spitzer}/IRAC [4.5] channel image \citep[][bottom]{ramirez:08}.  UKIDSS photometry plus our point-spread function fitting photometry shows that $L'$ = 9.3 for SSTGC~726327E, with $H-K=2.7$ and $K-L'=1.6$. The $L'$ (3.8 $\mu$m) photometry is in good agreement with [3.6] = 9.3 for SSTGC~726327 \citep{ramirez:08}.  The extended emission we detect at $L'$, but not at $K$, roughly coincides with the 3\arcsec\ (0.12 pc) diameter \ion{H}{2} region \object{Sgr~B1~A} \citep[][dotted circle]{mehringer:92}, also known as \object{GPSR5 0.488-0.028} \citep{becker:94}.

The IRAC source SSTGC~726327 is a composite of flux from SSTGC~726327E, SSTGC~726327W, the \ion{H}{2} region, and the MYSO. The IRAC position, relative to SSTGC~726327E, is $0.9\arcsec$ west at $4.5\ \mu$m, but is located further away ($1.1\arcsec$) at 8.0 $\mu$m \citep{ramirez:08}. \citet{morales:16} find that, when a candidate YSO selected based on {\it Spitzer}/IRAC photometry is matched to a UKIDSS source by their spectral energy distributions, 94\% of the {\it Spitzer} and UKIDSS sources are separated by $\leq$~0.57\arcsec.  This suggests SSTGC~726327 and SSTGC~726327E are distinct sources with a projected separation, for a GC distance of $8$~kpc, of $6000$ to $10,000$~au.

\begin{figure*}
\centering
\includegraphics[scale=0.6]{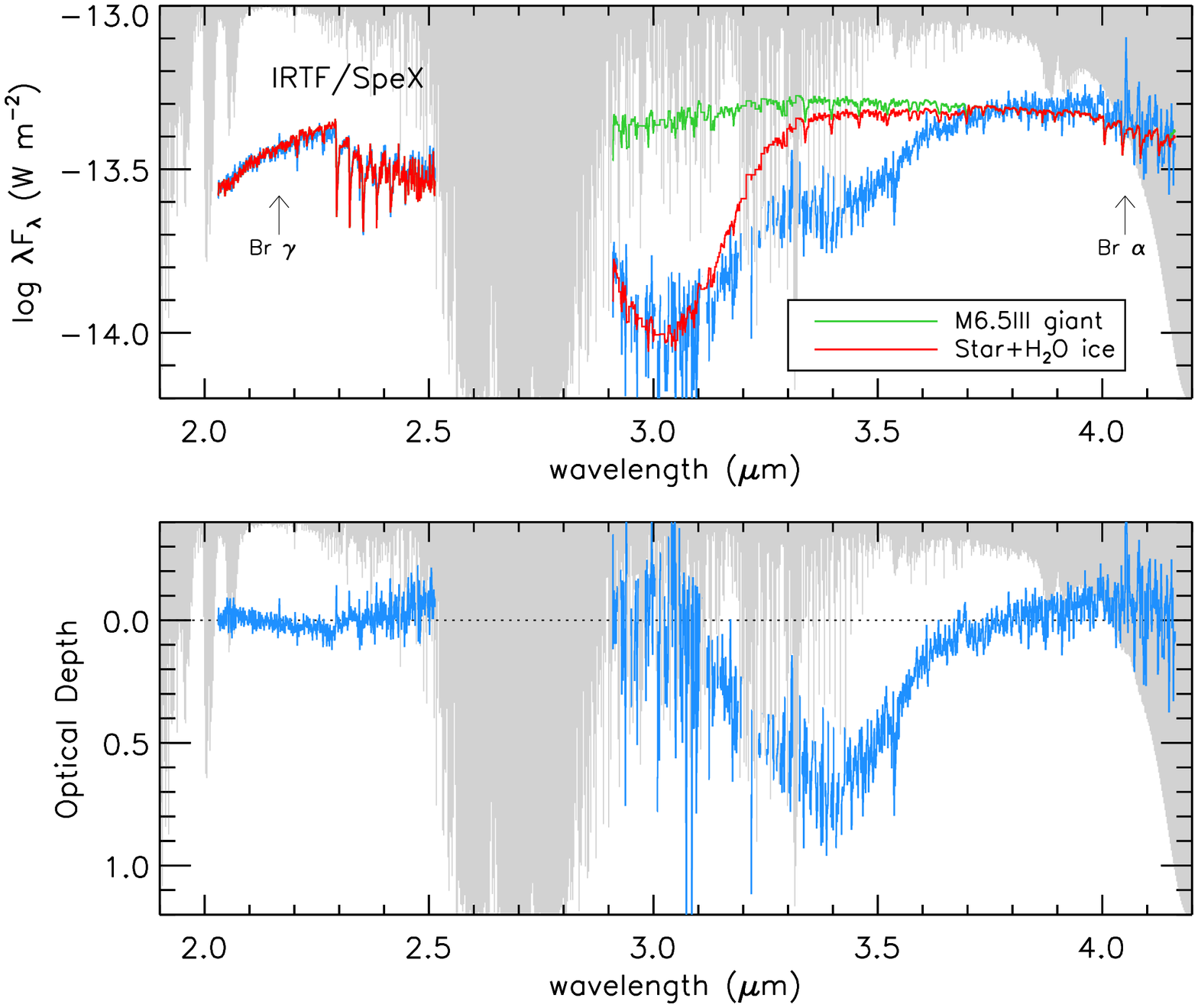}
\caption{Top: IRTF/SpeX spectra of SSTGC~726327E (blue line). The red line shows the best-fitting spectrum of the M6.5~III star (HD~94705) in the IRTF/SpeX library \citep{rayner:09} with added reddening and $3.0\ \mu$m H$_2$O ice absorption from $0.4\ \mu$m ice grains at $10$~K \citep{boogert:11}. The green line is the same best-fitting spectrum, but without H$_2$O ice absorption. Bottom: optical depth spectra computed using the best-fitting model. In each panel, theoretical atmospheric transmission curves at Mauna Kea are shown by gray shades scaled from $0$ to $1$.}\label{fig:irtfspectrum} \end{figure*}

The top panel in Figure~\ref{fig:irtfspectrum} shows our $2$--$4\ \mu$m IRTF/SpeX spectrum of SSTGC~726327E (blue line). The spectrum shows photospheric 2.3 $\mu$m CO band-head absorption, demonstrating that SSTGC~726327E is probably a background giant star rather than a MYSO. The observed Br~$\gamma$ ($2.166\ \mu$m) and Br~$\alpha$ ($4.051\ \mu$m) emission likely arise in the Sgr B1 A \ion{H}{2} region.

We searched for the best-fitting spectral type, the peak optical depth of the $3.0\ \mu$m H$_2$O ice band [$\tau_{\rm solid}$(H$_2$O)], and foreground dust extinction of SSTGC~726327E, by utilizing the IRTF/SpeX spectral library \citep{rayner:09}, the optical constants of $10$~K amorphous H$_2$O ice on $0.4\ \mu$m grains, and the extinction curve derived from the line of sight to a dense core \citep{boogert:11}. We minimized the reduced $\chi^2$ ($\chi^2_\nu$) of the fit to the SpeX spectra and $J$ and $H$ UKIDSS photometry ($J=18.391\pm0.068$ and $H=13.571\pm0.003$). We masked Br~$\alpha$ and Br~$\gamma$ emission lines, and excluded data points at $3.1\,\mu$m $\leq \lambda \leq 3.7\,\mu$m to avoid the $3.4\ \mu$m absorption band discussed later. We multiplied the $K$ and $L$ spectra by a constant factor ($1.16$) to match the UKIDSS $K$-band measurement ($K=10.915\pm0.003$).

We found that SSTGC~726327E is likely a M giant or K/M supergiant, along with $\tau_{\rm solid}$(H$_2$O)$=1.7\pm0.3$ and $A_K = 4.2\pm0.5$. The errors represent a range of values within $\Delta \chi^2_\nu < 2$ from the minimum $\chi^2_\nu$ ($0.83$). We found a $30\%$ smaller $A_K$ if we used the GC extinction curve from \citet{fritz:11} instead; spectral types and $\tau_{\rm solid}$(H$_2$O) essentially remain unchanged within the errors. The green line in the top panel of Figure~\ref{fig:irtfspectrum} shows our best-fitting spectrum (M6.5~III; \object{HD~142143}) with added reddening ($A_K=4.15$). The red line represents the best-fitting spectrum to our IRTF/SpeX data, with the $3\ \mu$m H$_2$O ice absorption [$\tau_{\rm solid}$(H$_2$O)$=1.51$] added to the reddened M6.5~III spectrum. The optical depth spectrum with respect to the best-fitting model is shown in the bottom panel of Figure~\ref{fig:irtfspectrum}. The strong, broad $3.4\ \mu$m absorption band in SSTGC~726327E is a blend of two distinct absorption features from icy molecular cloud grains \citep{brooke:99} and foreground diffuse cloud dust \citep{sandford:91,pendleton:94}.

\begin{figure*}
\centering
\includegraphics[scale=0.65]{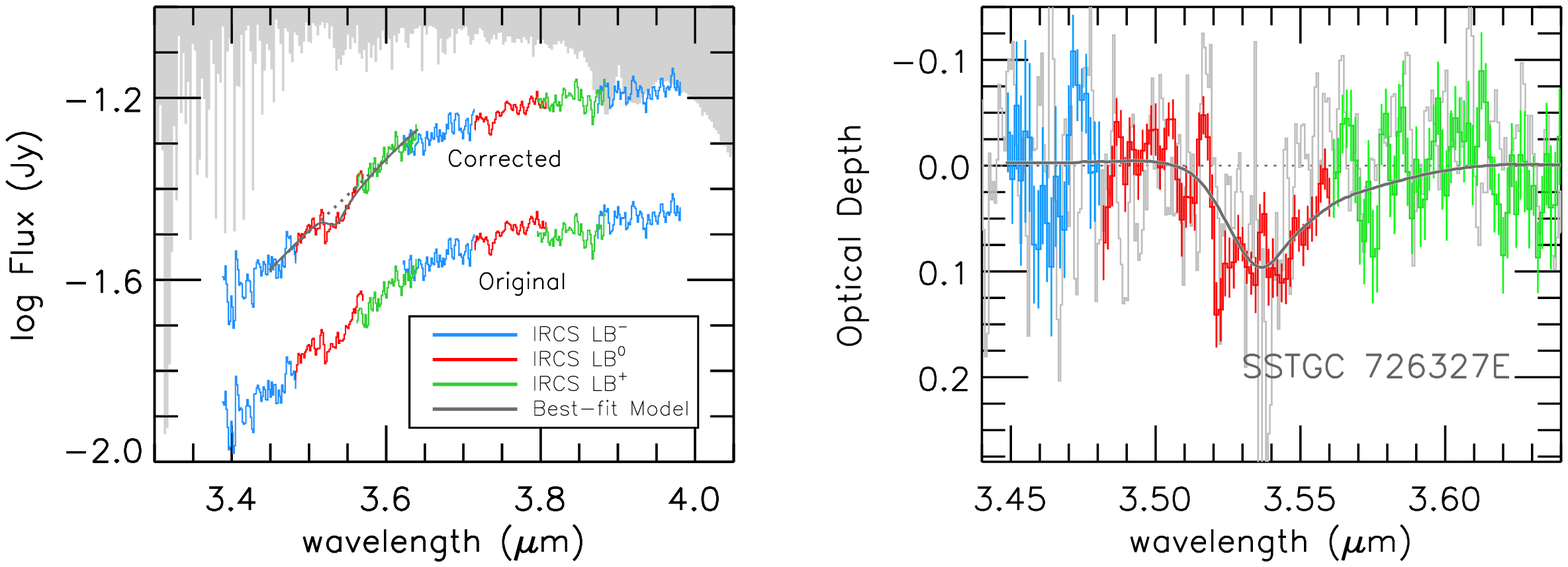}
\caption{Left: Subaru/IRCS $L$-band spectrum of SSTGC~726327E. The `original' spectra represent those calibrated and corrected for telluric absorptions. The `corrected' spectra are the same, except that additional zero-point flux offsets are applied in each of the three IRCS configurations (blue, red, and green lines for LB$^-$, LB$^0$, and LB$^+$, respectively) to match our IRTF/SpeX spectrum. The dotted line is a linear fit to the continuum around the $3.535\ \mu$m CH$_3$OH ice band, and the solid line represents the best-fitting laboratory spectrum of pure CH$_3$OH ice at $10$~K \citep{hudgins:93}. Theoretical atmospheric transmission curves at Mauna Kea are shown by grey shade scaled from $0$ to $1$. Right: optical depth spectra derived from the local continuum for SSTGC~726327E and the best-fitting model spectrum. IRTF/SpeX data are shown in gray line.} \label{fig:spectra} \end{figure*}

In the left panel of Figure~\ref{fig:spectra}, our $3.4$--$4.0\ \mu$m IRCS spectra of SSTGC~726327E are shown before (`original') and after scaling the flux to match the IRTF data (`corrected'). Methanol ice absorption is seen at $3.535\ \mu$m in our IRTF spectrum (Figure~\ref{fig:irtfspectrum}), but it is more clearly observed in the IRCS spectra. We do not detect the weaker CH$_3$OH features at $3.84\ \mu$m and $3.94\ \mu$m \citep{dartois:99b}.

We derived a local continuum across the $3.535\ \mu$m band using a first order polynomial fitted at $3.448 \leq \lambda \leq 3.497\ \mu$m ($2860$--$2900\ {\rm cm^{-1}}$) and $3.597 \leq \lambda \leq 3.650\ \mu$m ($2740$--$2780\ {\rm cm^{-1}}$) in wavenumber space (dotted line). The right panel in Figure~\ref{fig:spectra} shows the optical depth derived by dividing the `corrected' IRCS spectra by the local continuum. We employed a laboratory transmission spectrum of $10$~K pure CH$_3$OH ice \citep{hudgins:93}, in which the $3.535\ \mu$m band overlaps with another, wide band, centered at $3.39\ \mu$m ($2960$~cm$^{-1}$) from the $\nu_9$ C--H stretching mode. We fitted a first order polynomial to the laboratory spectrum over the same wavelength intervals adopted for the continuum construction, and isolated the $3.535\ \mu$m component by subtracting the local baseline. The solid lines in Figure~\ref{fig:spectra} show our best-fitting model spectrum for CH$_3$OH ice, which has a peak optical depth at $3.535\ \mu$m, $\tau_{\rm solid}$(CH$_3$OH), of $0.10\pm0.01$. This agrees with a result from an alternative approach, in which we restricted our fit to $3.45\ \mu$m--$3.64\ \mu$m and simultaneously searched for the flux scaling factors of individual IRCS orders and the best-fitting laboratory CH$_3$OH ice spectrum. 

\begin{deluxetable*}{llllrrr}
\tabletypesize{\scriptsize}
\tablecaption{Peak Optical Depth and Column Density of Solid H$_2$O and CH$_3$OH \label{tab:tab1} }
\tablehead{
  \colhead{Object(s)} &
  \colhead{$A_V$} &
  \colhead{$\tau_{\rm solid} (\rm H_2O)$} &
  \colhead{$N_{\rm solid} (\rm H_2O)$} &
  \colhead{$\tau_{\rm solid} (\rm CH_3OH)$} &
  \colhead{$N_{\rm solid} (\rm CH_3OH)$} &
  \colhead{$N_{\rm solid} ({\rm CH_3OH})/N_{\rm solid} ({\rm H_2O})$} \\
  \colhead{} &
  \colhead{(mag)} &
  \colhead{} &
  \colhead{$(10^{18}$ cm$^{-2})$} &
  \colhead{} &
  \colhead{$(10^{18}$ cm$^{-2})$} &
  \colhead{}
 }
\startdata
SSTGC~726327E & $24$--$43$\tablenotemark{a} & $1.7\pm0.3$ & $ 2.8\pm0.5$ & $0.10\pm0.01$ & $0.47\pm0.05$ & $0.17\pm0.03$ \\
\object{Sgr~A*} & $30$ & $0.50\pm0.01$ & $1.24\pm0.25$ & $<0.01$ & $<0.05$ & $<0.04$ \\
\objectname[WR 102dd]{GCS~3~I} & 29 & $0.23\pm0.02$ & $0.47\pm0.04$ & $<0.03$ & $<0.13$ & $<0.27$ \\
Massive YSOs & & & & & & $<0.03$ to 0.31 \\
Low-mass YSOs & & & & & & $<0.01$ to 0.25 \\
Quiescent clouds & & & & & & $<0.01$ to 0.12 \\
\enddata
\tablecomments{References for the Galactic center sources: SSTGC~726327E (this paper); Sgr~A* and GCS~3-I \citep{chiar:00,gibb:04}.  Minimum and maximum values for source classes from \citet{boogert:15}.}
\tablenotetext{a}{A range bracketed by values from our model fitting, either $A_K=4.2\pm0.5$ with a dense core extinction curve \citep{boogert:11} or a $30\%$ lower $A_K$ from model fitting with a GC extinction law \citep{fritz:11}. We assumed $A_K/A_V=0.11$.}
\end{deluxetable*}

Table~\ref{tab:tab1} summarizes the $3.535\ \mu$m CH$_3$OH ice optical depth and column density  for SSTGC~726327E. The shape and the intrinsic integrated band strength ($A$) for the $3.535\ \mu$m C--H stretch mode have a weak dependence on temperature and abundance of ice mantles \citep[e.g.,][]{ehrenfreund:99,kerkhof:99}. We averaged values from \citet{hudgins:93} and \citet{schutte:96} to adopt $A=5.95\pm0.65 \times10^{-18}$ cm molecule$^{-1}$ for CH$_3$OH ice, then derived its column density. We also list $\tau_{\rm solid}$(H$_2$O) for SSTGC~726327E. We computed $N_{\rm solid} (\rm H_2O)$ by multiplying $\tau_{\rm solid}$(H$_2$O) by FWHM $\approx330$~cm$^{-1}$, and assuming $A=2.0\times10^{-16}$~cm molecule$^{-1}$ \citep{hagen:81}.

\section{Discussion}

In this letter, we present the first detection of CH$_3$OH ice absorption toward the CMZ, in the background star SSTGC~726327E.  \citet{geballe:10} identified $2.27\ \mu$m absorption in a background GC star as possibly due to CH$_3$OH ice, but the absence of $3.53\ \mu$m CH$_3$OH ice absorption rules out this identification (T.\ R.\ Geballe, 2016, private communication). Table~\ref{tab:tab1} compares $N_{\rm solid}$(CH$_3$OH)/$N_{\rm solid}$(H$_2$O)$=0.17\pm0.03$ toward SSTGC 726327E to previous upper limits from a 14\arcsec$\times$20\arcsec\ aperture placed on Sgr A* and from the Quintuplet cluster star \objectname[WR 102dd]{GCS 3-I} \citep{chiar:00,gibb:04}. $N_{\rm solid}$(CH$_3$OH)/$N_{\rm solid}$(H$_2$O) is $\ga 4$ times higher toward SSTGC~726327E than toward Sgr~A*.

SSTGC~726327E is $\sim8000$~au in projection from the MYSO SSTGC~726327 in Sgr~B1. The CH$_3$OH ice absorption we detect toward SSTGC~726327E likely arises from the extended envelope of the MYSO, which can extend to $\sim30,000$--$90,000$~au \citep[e.g.,][]{vandertak:00b}. \citet{pontoppidan:04} have measured column densities of CH$_3$OH ice from an extended envelope of a low-mass Class~0 protostar, \objectname[NAME Serpens SMM 4]{Serpens SMM~4}, based on the $L-$band spectra of 10 pre-main sequence stars. These `background' stars show a constant $N_{\rm solid}$ $(\rm CH_3OH)$/$N_{\rm solid}$ $(\rm H_2O)\sim0.3$ at projected distances of $4000$--$10,000$~au from Serpens~SMM~4, then no CH$_3$OH at a projected distance of $19,000$~au. They conclude, as we do, that some or all CH$_3$OH ice absorption arises from the extended envelope of the YSO.

While we have discovered CH$_3$OH ice in one line of sight to CMZ, CH$_3$OH ice is found in abundance not only in YSOs, but also toward stars behind cold, quiescent molecular clouds and cores in the Galactic disk at $A_V\ga9$~mag \citep[Table~\ref{tab:tab1};][]{boogert:15}. For two disk MYSOs \citep[RAFGL~7009S and W33~A;][]{brooke:99,dartois:99b}, $N_{\rm solid}$(CH$_3$OH)/$N_{\rm solid}$(H$_2$O) is higher than the maximum value observed toward stars behind quiescent cores in the disk \citep[0.12;][]{boogert:11}. This empirical division also makes the case for SSTGC~726327E being behind the envelope of the MYSO SSTGC~726327.

If SSTGC~726327 has physical and chemical structures similar to those of W33~A \citep{vandertak:00a,vandertak:00b}, the maximum gas density encountered along the line of sight at the projected distance of $8000$~au is $\sim6\times10^5$~cm$^{-3}$, which is sufficiently large enough to maintain CH$_3$OH formation in its envelope. However, the gas temperature within the MYSO envelope likely exceeds the sublimation temperature of CO ($\sim20$~K) all the way to the edge of the MYSO, under which additional growth of CH$_3$OH on ice mantles should be suppressed \citep{watanabe:03,cuppen:09}. CMZ clouds are on average warmer than in the disk, but there are dense shielded regions with cold ($\sim15$~K) dust grains \citep{rodriguez:04,molinari:11}, where CH$_3$OH can form. Then it is possible that CH$_3$OH ice grains arise in dense molecular clouds within the CMZ, in which case the star SSTGC~726327E is projected by chance next to the MYSO.

The fractional abundance of CH$_3$OH ice with respect to H$_2$ is $\sim10^{-5}$ toward SSTGC~726327E, if our best-fitting $A_K$ is taken with $A_K/A_V=0.11$ and $N({\rm H_2})/A_V \sim 10^{21}$~cm$^{-2}$~mag$^{-1}$ \citep[see][and references therein]{hasenberger:16}. This order of magnitude estimate is $\sim10^1$--$10^3$ times larger than the gas-phase CH$_3$OH abundance in the CMZ \citep{requena-torres:06,yusefzadeh:13}. While systematic searches for CH$_3$OH ices should be proceeded before making any firm conclusions, our estimate suggests that gas-phase CH$_3$OH in the CMZ can be largely produced by desorption of CH$_3$OH from icy grains \citep[e.g.,][]{requena-torres:08,yusefzadeh:13,coutens:17}.

In spite of abundant CH$_3$OH ices in various sight lines, broad CO$_2$ ice bands with the $15.4\ \mu$m shoulder absorption have, to date, only been observed toward YSOs \citep{boogert:15}. About $19\%$ of the CO$_2$ absorption in SSTGC~726327 is attributed to the shoulder CO$_2$ \citep{an:11}. With our measurement of CH$_3$OH abundance, we found $N_{\rm solid}$(CH$_3$OH)/$N_{\rm solid}$(CO$_2$ shoulder)$=3.9\pm0.4$. We applied the same spectral decomposition procedure to the observed CO$_2$ ice profile of the ISO spectrum of W33~A \citep{gerakines:99}, and found $10\%$ for the fraction of the shoulder component, which results in $N_{\rm solid}$(CH$_3$OH)/$N_{\rm solid}$(CO$_2$ shoulder)$=11.2\pm2.4$ based on $N_{\rm solid}$(CH$_3$OH) in \citet{brooke:99}. Although there exists a factor of three difference in this comparison, systematic errors likely dominate, since our measurement only traces CH$_3$OH ice in the outer part of the MYSO. The similar $N_{\rm solid}$(CH$_3$OH)/$N_{\rm solid}$(CO$_2$ shoulder) implies that CH$_3$OH is indeed the best candidate for interacting molecules in the observed $15.4\ \mu$m shoulder CO$_2$ toward GC MYSOs \citep{an:09,an:11}, as it is in disk MYSOs. Additional observations are clearly needed to extend a source list, establish their properties, and estimate the variance of CH$_3$OH ice in the CMZ.

\acknowledgements

D.A.\ acknowledges support provided by Basic Science Research Program through the National Research Foundation of Korea (NRF) funded by the Ministry of Education (NRF-2015R1D1A1A09058700) and by the Korean NRF to the Center for Galaxy Evolution Research (No.\ 2010-0027910).

Facilities: \facility{IRTF}, \facility{Subaru}

{}


\begin{thebibliography}{}

\bibitem[An et al.(2009)]{an:09} An, D., Ram{\'{\i}}rez, S.~V., Sellgren, K., et al.\ 2009, \apjl, 702, L128 

\bibitem[An et al.(2011)]{an:11} An, D., Ram{\'{\i}}rez, S.~V., Sellgren, K., et al.\ 2011, \apj, 736, 133 

\bibitem[Becker et al.(1994)]{becker:94} Becker, R.~H., White, R.~L., Helfand, D.~J., \& Zoonematkermani, S.\ 1994, \apjs, 91, 347 

\bibitem[Boogert et al.(2015)]{boogert:15} Boogert, A.~C.~A., Gerakines, P.~A., \& Whittet, D.~C.~B.\ 2015, \araa, 53, 541

\bibitem[Boogert et al.(2011)]{boogert:11} Boogert, A.~C.~A., Huard, T.~L., Cook, A.~M., et al.\ 2011, \apj, 729, 92 

\bibitem[Brooke et al.(1999)]{brooke:99} Brooke, T.~Y., Sellgren, K., \& Geballe, T.~R.\ 1999, \apj, 517, 883 

\bibitem[Castelli \& Kurucz(2004)]{castelli:04} Castelli, F., \& Kurucz, R.~L.\ 2004, arXiv:astro-ph/0405087 

\bibitem[Chiar et al.(2000)]{chiar:00} Chiar, J. E., Tielens, A. G. G. M., Whittet, D. C. B., Schutte, W. A., Boogert, A. C. A., Lutz, D., van Dishoeck, E. F., \& Bernstein, M. P. 2000, ApJ, 537, 749

\bibitem[Coutens et al.(2017)]{coutens:17} Coutens, A., Rawlings, J.~M.~C., Viti, S., \& Williams, D.~A.\ 2017, \mnras, 467, 737

\bibitem[Cuppen et al.(2009)]{cuppen:09} Cuppen, H.~M., van Dishoeck, E.~F., Herbst, E., \& Tielens, A.~G.~G.~M.\ 2009, \aap, 508, 275 

\bibitem[Cushing et al.(2004)]{cushing:04} Cushing, M.~C., Vacca, W.~D., \& Rayner, J.~T.\ 2004, \pasp, 116, 362 

\bibitem[Dartois et al.(1999a)]{dartois:99a} Dartois, E., Demyk, K., d'Hendecourt, L., \& Ehrenfreund, P.\ 1999a, \aap, 351, 1066 

\bibitem[Dartois et al.(1999b)]{dartois:99b} Dartois, E., Schutte, W., Geballe, T.~R., et al.\ 1999b, \aap, 342, L32 

\bibitem[Ehrenfreund et al.(1999)]{ehrenfreund:99} Ehrenfreund, P., Kerkhof, O., Schutte, W.~A., et al.\ 1999, \aap, 350, 240 

\bibitem[Fritz et al.(2011)]{fritz:11} Fritz, T. K.,  Gillessen, S., Dodds-Eden, K., Lutz, D., Genzel, R., Raab, W., Ott, T., Pfuhl, O., Eisenhauer, F., \& Yusef-Zadeh, F.  2011, ApJ, 737, 73

\bibitem[Geballe \& Oka(2010)]{geballe:10} Geballe, T. R., \& Oka, T. 2010, ApJL, 209, L70

\bibitem[Gerakines et al.(1999)]{gerakines:99} Gerakines, P.~A., Whittet, D.~C.~B., Ehrenfreund, P., et al.\ 1999, \apj, 522, 357

\bibitem[Gibb et al.(2004)]{gibb:04} Gibb, E.~L., Whittet, D.~C.~B., Boogert, A.~C.~A., \& Tielens, A.~G.~G.~M.\ 2004, \apjs, 151, 35

\bibitem[Goto et al.(2008)]{goto:08} Goto, M., Usuda, T., Nagata, T., Geballe, T. R., McCall, B. J., Indriolo, N., Suto, H., Henning, T.,  Morong, C. P., \& Oka, T.  2008, ApJ, 688, 306

\bibitem[Hagen et al.(1981)]{hagen:81} Hagen, W., Tielens, A.~G.~G.~M., \& Greenberg, J.~M.\ 1981, Chemical Physics, 56, 367 

\bibitem[Hasenberger et al.(2016)]{hasenberger:16} Hasenberger, B., Forbrich, J., Alves, J., et al.\ 2016, \aap, 593, A7

\bibitem[Hudgins et al.(1993)]{hudgins:93} Hudgins, D.~M., Sandford, S.~A., Allamandola, L.~J., \& Tielens, A.~G.~G.~M.\ 1993, \apjs, 86, 713 

\bibitem[Kerkhof et al.(1999)]{kerkhof:99} Kerkhof, O., Schutte, W.~A., \& Ehrenfreund, P.\ 1999, \aap, 346, 990 

\bibitem[Kobayashi et al.(2000)]{kobayashi:00} Kobayashi, N., Tokunaga, A.~T., Terada, H., et al.\ 2000, \procspie, 4008, 1056 

\bibitem[Lucas et al.(2008)]{lucas:08} Lucas, P.~W., Hoare, M.~G., Longmore, A., et al.\ 2008, \mnras, 391, 136 

\bibitem[Moultaka et al.(2015)]{moultaka:15} Moultaka, J., Eckart, A., \& Mu{\v z}i{\'c}, K.\ 2015, \apj, 806, 202 

\bibitem[Mehringer et al.(1992)]{mehringer:92} Mehringer, D.~M., Yusef-Zadeh, F., Palmer, P., \& Goss, W.~M.\ 1992, \apj, 401, 168 

\bibitem[Molinari et al.(2011)]{molinari:11} Molinari, S., Bally, J., Noriega-Crespo, A., et al.\ 2011, \apjl, 735, L33 

\bibitem[Morales et al.(2016)]{morales:16} Morales, E.~F.~E., \& Robitaille, T.~P.\ 2017, \aap, 598, A136

\bibitem[Morris \& Serabyn(1996)]{morris:96} Morris, M., \& Serabyn, E.\ 1996, \araa, 34, 645 

\bibitem[Pendleton et al.(1994)]{pendleton:94} Pendleton, Y.~J., Sandford, S.~A., Allamandola, L.~J., Tielens, A.~G.~G.~M., \& Sellgren, K.\ 1994, \apj, 437, 683

\bibitem[Pontoppidan et al.(2004)]{pontoppidan:04} Pontoppidan, K.M., van Dishoeck, E.F., \& Dartois,  E. 2004, A\&A, 426, 925

\bibitem[Ram{\'{\i}}rez et al.(2008)]{ramirez:08} Ram{\'{\i}}rez, S.~V., Arendt, R.~G., Sellgren, K., et al.\ 2008, \apjs, 175, 147

\bibitem[Rawlings et al.(2013)]{rawlings:13} Rawlings, J.~M.~C., Williams, D.~A., Viti, S., Cecchi-Pestellini, C., \& Duley, W.~W.\ 2013, \mnras, 430, 264 

\bibitem[Rayner et al.(2003)]{rayner:03} Rayner, J.~T., Toomey, D.~W., Onaka, P.~M., et al.\ 2003, \pasp, 115, 362 

\bibitem[Rayner et al.(2009)]{rayner:09} Rayner, J.~T., Cushing, M. C., \& Vacca, William D.  
2009, ApJS, 185, 289

\bibitem[Requena-Torres et al.(2006)]{requena-torres:06} Requena-Torres, M. A., Mart{\'{\i}}n-Pintado, J., Rodr{\'{\i}}guez-Franco, A., Mart{\'{\i}}n, S., Rodr{\'{\i}}guez-Fern{\'a}ndez, N. J., \& de Vicente, P.  2006, A\&A, 455, 971

\bibitem[Requena-Torres et al.(2008)]{requena-torres:08} Requena-Torres, M. A., Mart{\'{\i}}n-Pintado, J., Mart{\'{\i}}n, S., \& Morris, M. R. 2008, ApJ, 672, 352 

\bibitem[Rodr{\'{\i}}guez-Fern{\'a}ndez et al.(2004)]{rodriguez:04} Rodr{\'{\i}}guez-Fern{\'a}ndez, N.~J., Mart{\'{\i}}n-Pintado, J., Fuente, A., \& Wilson, T.~L.\ 2004, \aap, 427, 217 

\bibitem[Sandford et al.(1991)]{sandford:91} Sandford, S.~A., Allamandola, L.~J., Tielens, A.~G.~G.~M., et al.\ 1991, \apj, 371, 607

\bibitem[Schultheis et al.(2009)]{schultheis:09} Schultheis, M., Sellgren, K., Ram{\'{\i}}rez, S., Stolovy, S., Ganesh, S., Glass, I. S., \& Girardi, L. 2009, A\&A, 495, 157

\bibitem[Schutte et al.(1996)]{schutte:96} Schutte, W.~A., Gerakines, P.~A., Geballe, T.~R., van Dishoeck, E.~F., \& Greenberg, J.~M.\ 1996, \aap, 309, 633

\bibitem[Shure et al.(1994)]{shure:94} Shure, M.~A., Toomey, D.~W., Rayner, J.~T., Onaka, P.~M., \& Denault, A.~J.\ 1994, \procspie, 2198, 614 

\bibitem[Vacca et al.(2003)]{vacca:03} Vacca, W.~D., Cushing, M.~C., \& Rayner, J.~T.\ 2003, \pasp, 115, 389 

\bibitem[van der Tak et al.(2000a)]{vandertak:00a} van der Tak, F.~F.~S., van Dishoeck, E.~F., \& Caselli, P.\ 2000a, \aap, 361, 327 

\bibitem[van der Tak et al.(2000b)]{vandertak:00b} van der Tak, F.~F.~S., van Dishoeck, E.~F., Evans, N.~J., II, \& Blake, G.~A.\ 2000b, \apj, 537, 283 

\bibitem[Watanabe et al.(2003)]{watanabe:03} Watanabe, N., Shiraki, T., \& Kouchi, A.\ 2003, \apjl, 588, L121

\bibitem[Yusef-Zadeh et al.(2013)]{yusefzadeh:13} Yusef-Zadeh, F., Cotton, W., Viti, S., Wardle, M., \& Royster, M.\ 2013, \apjl, 764, L19 

\end{thebibliography}
\end{document}